\documentclass[11pt]{article}
\usepackage{appb,epsfig}
\input psfig

\def\xslide#1#2#3#4#5#6#7{\centerline{\psfig
{figure=#1,height=#2,bbllx=#3bp,bblly=#4bp,bburx=#5bp,bbury=#6bp,width=#7,clip=}}}

\begin{document}


\newcommand{\kk}{$K\overline{K}$ }
\newcommand{\kpkm}{$K^+K^-$ }
\newcommand{\pp}{$\pi^+\pi^-$ }

\title{Photoproduction of ${\pi}{\pi}$ and \kk Pairs}

\author{Leonard LE\'SNIAK
\address{Department of Theoretical Physics,
H. Niewodnicza{\'n}ski Institute of Nuclear Physics,
PL 31-342~ Krak\'ow, Poland}
}
\maketitle

\begin{abstract}

S-wave photoproduction of  $\pi^+\pi^-$ and \kpkm pairs on hydrogen is studied.
Coupled channel final state interactions between different pseudoscalar mesons
are included. Effective mass and momentum transfer distributions as well as 
total cross sections are calculated. Effective  mass distributions in the Born
approximation are structureless but the final state interactions produce peaks
and dips near the energies corresponding to the scalar re\-so\-nances $f_0$(980) 
and $f_0$(1400). Calculations are performed in the effective mass region 
between the $\pi\pi$ threshold and 1.6 GeV in the range of the momentum transfer squared 
up to 2 (GeV/c)$^2$. The photon energy dependence of the total cross 
sections is studied between 2 and 15 GeV. Valuable information on the $f_0$(980) resonance
structure can be extracted from data.

   These calculations are closely related to the experimental program of the 
   CEBAF accelerator in USA.

\end{abstract}

\vspace{6mm} 

Properties of scalar
mesons can be studied in photoproduction and electroproduction processes. 
Asymmetry in the $K^+ K^-$ 
angular distribution near threshold was previously measured at Daresbury 
\cite{barberetal} and Hamburg \cite{Friesetal} and interpreted as an 
interference of the dominant $P$-wave from the decay of the $\phi(1020)$ meson 
and an $S$-wave from the decay of the $f_0(980)$ resonance. The 
existing data are imprecise and even controversial. This si\-tuation may be
improved in near future when a new program of the  $f_0(980)$ electroproduction
experiments on the CEBAF accelerator at the Thomas Jefferson National
Accelerator Facility (TJNAF) is realized. 
Therefore we have performed a coupled channel analysis of the
$S$-wave $\pi^+\pi^-$
and \kpkm photoproduction on hydrogen \cite{kolab}. 

We expect that at high photon energies 
($E_{\gamma} \geq 4 \;$ GeV) mostly the t-channel $\rho$ and
$\omega$ exchanges become important in the peripheral $\pi^+\pi^-$ and
\kpkm production. Five Feynman diagrams are considered to calculate the Born 
amplitude for the \pp production, three of them correspond to $\pi^{\pm}$ and
$\rho^0$ exchange and two to $\rho^{\pm}$ and $\omega$ exchange. For the six
diagrams of $K^+ K^-$ production, three of them correspond to $K^{\pm}$ and 
$\rho^0$ exchange and three to $K^{\pm}$ and $\omega$ exchange. We require 8 
hadronic ($g_{\rho \pi \pi}$, $g_{\rho K K}$, $g_{\omega K K}$,
 $g_{\omega \rho \pi}$, $G_{V}^{\rho}$, $G_{T}^{\rho}$, $G_{V}^{\omega}$,
 $G_{T}^{\omega}$) and 2 electromagnetic ($g_{\rho \pi \gamma}$,
 $g_{\omega \pi \gamma}$) vector meson coupling constants. 
From the $\rho \rightarrow \pi \pi$ decay width, we find the
coupling constant $g_{\rho \pi \pi} = 6.05$. We use the SU(3) relations
$g_{\rho K {\bar K}} = g_{\omega K {\bar K}} = \frac{1}{2} g_{\rho \pi \pi}$ 
to fix the kaon couplings and $g_{\omega \rho \pi} = 14.0$ GeV$^{-1}$,
  close to the value reported
in Ref.~\cite{Bramon}. For the $\rho$ meson vector and tensor couplings to 
nucleon, the values corresponding to the Bonn potential  
($G_{V}^{\rho} = 2.27$, $G_{T}^{\rho} = 13.85$,$G_{T}^{\omega} = 0$ and
$G_{V}^{\omega} = 11.54$) are applied ~\cite{Machleidt}. Finally, we fit the 
radiative decay constants of the
$\rho$ and $\omega$ to the $\Gamma_{\rho \rightarrow \pi \gamma}$ and
$\Gamma_{\omega \rightarrow \pi \gamma}$ decay widths
yielding $g_{\rho \pi \gamma} = 0.75e/m_{\rho}$ and 
$g_{\omega \pi \gamma} = 1.82e/m_{\omega}$ with $e= 0.30282$.

The full photoproduction amplitude including final state interactions (FSI) can 
be written in the operator form as 

\begin{equation}
{\hat T} = {\hat V} + 
{\hat t} \hat G {\hat V}, \label {FFSI}
\end{equation}
where ${\hat V}$ is the photoproduction potential, $\hat t$ is the strong 
FSI matrix, and $\hat G$ is the propagator of the intermediate state. 
The matrix elements of $\hat V$ are obtained through an off-shell extension 
of the Born amplitudes.

As shown in references \cite{klm,kl}, several intermediate
 channels can contribute to the production of a given final meson pair.
Thus, the final state interaction leads to a coupled channel problem
for  $\hat T $ in Eq.(\ref{FFSI}). 
The isospin decomposition of each final state requires the inclusion of 
other possible meson pairs such as $\pi^0 \pi^0$ and $K^0 {\bar K}^0$
and thus one has to consider all four channels ($\pi^0 \pi^0$,
$\pi^+ \pi^-$, $K^+ K^-$ and $K^0 {\bar K}^0$) as the intermediate states. 
Explicit expressions for the $I=0$ matrix elements can
be found in the Appendix A of Ref.~\cite{klm}. 
Parameterization of the $I=2$ elastic $\pi \pi$ amplitude is given in 
Ref.~\cite{klr}. For the $I=0$ \pp amplitude we use parameters obtained
in a recent analysis of the
 $\pi^- p_{\uparrow} \rightarrow \pi^+ \pi^- n$ reaction on polarized target
\cite{kll}. They correspond to the two--channel fit to the so--called 
"down--flat" \pp phase shift solution of Ref. \cite{klr}.

In Fig.1 we show the $S$-wave $\pi^{+} \pi^{-}$ invariant mass distribution at 
$E_{\gamma}^{lab} = 4.0 \;$ GeV and $t = -0.2 \;$ GeV$^2$. 
\begin{figure}

\xslide{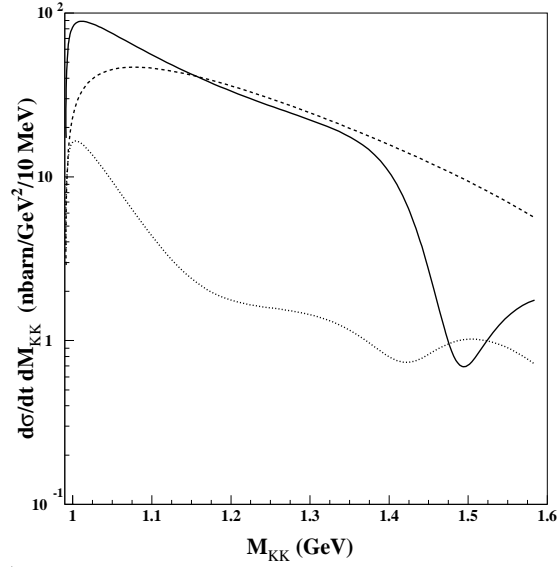}{7.5cm}{10}{145}{530}{655}{7.5cm} 

\caption{$S$-wave $\pi^{+} \pi^{-}$ invariant mass distribution.
 The solid line 
shows the full FSI result (on--shell and off--shell) with both $\pi \pi$ and 
$K\bar{K}$ intermediate channels, the dotted line represents the result with
no $K\bar{K}$ coupling and the dashed line corresponds to the Born cross 
section.}
\vspace{0.5cm}

\label{fig1}
\end{figure}
The Born effective mass distributions are structureless
while the final state interactions produce dips and peaks near the resonances.
In particular near 1 GeV we can notice a very clear maximum corresponding
to the $f_0(980)$ resonance. This signal should be expe\-ri\-mentally measurable in 
the photoproduction process at the photon energies of a few GeV. In Fig. 2
 we show the corresponding $K^{+} K^{-}$ invariant mass distribution.
\begin{figure}

\xslide{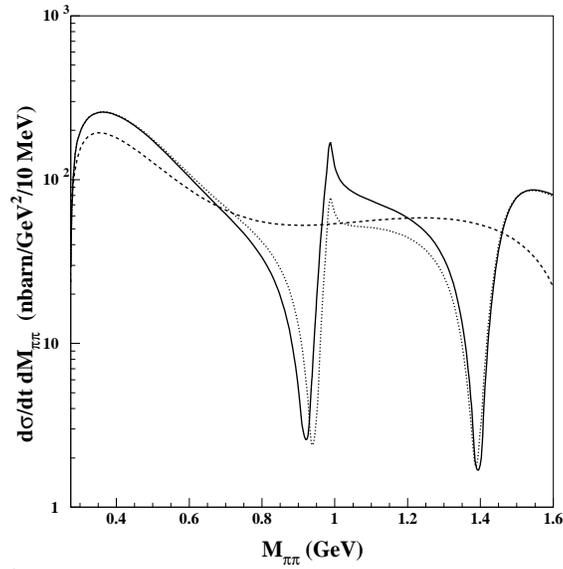}{7.5cm}{10}{145}{530}{655}{7.5cm} 

\caption{$S$-wave $K^{+} K^{-}$ invariant mass distribution. 
The dotted line represents the result with no $\pi\pi$ coupling, for
description of other lines see Fig. 1.}


\label{fig2}
\end{figure}
We see that the coupled channel FSI give a substantial enhancement
relative to the Born cross section just above the $K\bar{K}$
threshold that is absent in the result for the single channel FSI. 
This enhancement of the cross section in two kaon and two pion photoproduction
at the $f_{0}(980)$ resonance makes direct measurements
of $f_{0}$ properties an interesting possibility at TJNAF.
The differential cross section for the $S$-wave two kaon or two pion 
photoproduction has a dip at t-values close to 0 (forward production) and then
rises reaching a maximum at higher values of t. Calculations performed with Regge
 $\rho$ and $\omega$ propagators show a characteristic minimum at 
 $t \approx -0.5$ GeV$^2$ related to the zeroes of the Regge trajectories.
  In order to see the interference effect between 
 $S$ and $P$ wave contributions,
one should not limit the range of $|t|$ below 0.2 GeV$^2$ as in 
Ref.\cite{Friesetal} but go up to higher values such as 1.5 GeV$^2$ in
\cite{barberetal}.   
    
Our predictions of the total cross sections indicate that the $\pi^+\pi^-$ 
and $K^+K^-$
photoproduction processes are experimentally measurable in the photon energy
range of a few GeV. The S-wave $K^+K^-$ cross section near 1 GeV is, however, 
lower than the $\Phi$ production cross section.
Nethertheless, by a careful
experimental study in small bins of $\pi \pi$ and $K {\bar K}$ masses,
one can obtain valuable information on the positions and widths of scalar
resonances, especially the $f_0(980)$ meson.
Additionally, the accurate measurements of the neutral pion pairs
$\pi^0 \pi^0$ and $K^0 {\bar K}^0$ photoproduction will be crucial to
separate the different isospin contributions of $I=0,1$ and 2 states.\\
--------------------------------------------------------------------------------------------------\\
{\em $^*$ Presented at MESON'98 Workshop on Production, Properties and 
Interaction of Mesons, Cracow, Poland, May 30 - June 2, 1998, proceedings 
will be published in Acta Physica Polonica B.}

\end{document}